# Impulse Noise and Narrowband PLC

A.J. Han Vinck[(1,4)], F. Rouissi[(2)], T. Shongwe[(4)], G.R. Colen[(3)] and L.G. Oliveira[(3)]

[(1)] University Duisburg-Essen, Germany, email: han.vinck@uni-due.de

[(2)] University Tunis, Tunisia, email: rouissi_fatma@yahoo.com

[(3)] University Juiz de Fora, Brazil, email: lucas.giroto@engenharia.ufjf.br, guilherme.colen@engenharia.ufjf.br

[(4)] University of Johannesburg, Johannesburg, South Africa, email: caltoxs@gmail.com

**Abstract**

We discuss the influence of random- and periodic impulse noise on narrowband (< 500 kHz frequency band) Power Line Communications. We start with random impulse noise and compare the properties of the measured impulse noise with the common theoretical models like Middleton Class-A and Mixed Gaussian. The main difference is the fact that the measured impulse noise is noise with memory for the narrowband communication, whereas the theoretical models are memoryless. Since the FFT can be seen as a randomizing, operation, the impulse noise is assumed to appear as Gaussian noise after the FFT operation with a variance that is determined by the energy of the impulses. We investigate the problem of capacity loss due to this FFT operation. Another topic is that of periodical noise. Since periodic in the time domain means periodic in the frequency domain, this type of noise directly influences the output of the FFT for an OFDM based transmission. Randomization is necessary to avoid bursty- or dependent errors

## I. INTRODUCTION

Impulse noise is an important disturbance in Power Line Communications (PLC). Different types of noise are classified by Zimmermann, Dostert and Cortes et al. [2,8]. Two theoretical memoryless models dominate the literature: Middleton Class-A and Mixed Gaussian. Both models use the parameters A (for impulse frequency) and the variances of the impulse – and background noise. The 2-state Mixed Gaussian model can be seen as a simplification of the Middleton Class-A model. OFDM is a popular transmission method in PLC to overcome many different channel disturbances. OFMD uses the IFFT to transform input symbols to channel symbols. The FFT is used at the receiver. The applied transmission parameters for the CENELEC - and FCC band are given in Table 1.

Table 1
**OFDM parameters for the CENELEC and FCC band**

| Parameters | CENELEC band | FCC band |
|---|---|---|
| Frequency band | [35.2 KHz, 91.4 KHz] | [152.3, 489.8 KHz] |
| FFT (used subcarriers) | 256 (72) | 256 (72) |
| Time OFDM frame | 695 μs | 231,7 μs |
| Sample duration | 2.5 μs | 0.833 μs |
| Sampling frequency | 400 KHz | 1.2 MHz |
| Max. bitrate | 33.4 Kbit/s | 303 Kbit/s |

In our analysis, we assume that the FFT at the receiver side fully randomizes the impulse noise over an OFDM frame. As a result, the transmission is equivalent to transmission over a channel with two independent additive Gaussian noise sources. The first noise source is the Gaussian background noise and the second noise source is Gaussian with variance determined by the randomized impulse energy over an OFDM frame. For this memoryless channel we can determine the channel capacity. In Section II we consider the channel capacity for the situations where the transmitter and receiver have information about the channel state of the Mixed Gaussian channel. From a measurement campaign we obtained practical values for the parameters of the impulse noise: Pulse duration; Inter-arrival time and Power Spectral Density (PSD) per pulse. These values are used to estimate the variance of the noise in an OFDM frame for CENELEC - and FCC band. From the obtained variance per carrier we can determine whether impulse noise is a serious disturbance for the OFDM based PLC system, see Section III. Section IV gives an overview of noise mitigation techniques.

The last part of our contribution, Section V, discusses the effect of periodic noise. Periodic noise can be generated by switches, motors, light dimmers, etc. Since the FFT in the OFDM receiver transforms the periodic time domain samples into periodic frequency domain disturbances, this type of noise can be catastrophic for the applied error correcting tools.

## II. CAPACITY OF THE IMPULSE NOISE CHANNEL

The capacity of the impulse noise channel can be calculated for certain special situations [9]. It is the problem to determine the input distribution that maximizes the amount of transmitted information. We consider three situations:

1. *The fully randomized channel*. In this case, the channel is a purely Gaussian channel with noise variance $\sigma^2_G + \sigma^2_I$, where $\sigma^2_G$ is the background noise variance and $\sigma^2_I$ the impulse noise variance. For a maximum average Gaussian input power P, the channel capacity for a channel bandwidth B, where $P/2B := E_b$, is given by

$$C_1 = B\log_2(1 + \frac{E_b}{\sigma^2_G + \sigma^2_I}) \text{bit/s}.$$

2. *Receiver knows the channel state, transmitter does not*. Using the Mixed Gaussian model, the state with impulse noise has probability A and variance $(\sigma^2_G + \sigma^2_I/A)$. The transmitter does not know the state and uses a Gaussian input distribution with maximum average power P. For these conditions,

$$C_2 = (1-A)B\log_2(1 + \frac{E_b}{\sigma^2_G}) + AB\log_2(1 + \frac{E_b}{\sigma^2_G + \sigma^2_I/A}) \text{bit/s}.$$

3. *Receiver and transmitter know the channel state*: We note that this is not a practical situation. However, this situation can be used as an upper bound on the capacity. Using the water-filling argument in the time domain, the capacity (for large enough power) is given by:

$$C_3 = B\log_2(1 + \frac{E_b + \sigma^2_I}{\sigma^2_G}) + AB\log_2(\frac{\sigma^2_G}{\sigma^2_G + \sigma^2_I/A}) \text{bit/s}.$$

For large $E_b$ the difference between $C_1$ and $C_2$ is determined by $(\sigma^2_I + \sigma^2_G)/\sigma^2_G$. Hence, one should try to include state estimation at the receiver side to improve the performance. As an example, we use the following values (see Table 1-3):

FCC-Band: B = 337 kHz, A = 0.3, $E_b$ = 5.6 x $10^{-9}$ J; $\sigma^2_G$ = 5.6 x $10^{-16}$ J (-120dBm, SNR = 30 dB), $\sigma^2_I$ = 1.4 x $10^{-10}$ J.
We have: $C_1$ = 1.8 Mbit/s; $C_2$ = 7.8 Mbit/s; $C_3$ = 7.8 Mbit/s. We observe, that $C_2 \approx C_3 \approx$ 5 x $C_1$.
CENELEC-Band: B = 56.2 kHz; A = 0.3; $E_b$ = 2.8 $10^{-9}$ J; $\sigma^2_G$ = 2.8 $10^{-12}$ J(-80 dBm, SNR = 70dB); $\sigma^2_I$ = 1.4 x $10^{-10}$ J.
We have: $C_1$ = 240 kbit/s; $C_2$ = 500 kbit/s; $C_3$ = 560 kbit/s. We observe, that $C_2 \approx C_3 \approx$ 2 x $C_1$.

Table 2
**Summary of the performance of the OFDM modulation for the measured data**

| Parameters | Average | Highest probability | Worst (10%) | Best( 10%) |
|---|---|---|---|---|
| CENELEC | | | | |
| $\sigma^2_I$ | 5.3 $10^{-11}$ J | 1.7 $10^{-11}$ J | 18 $10^{-11}$ J | 3.3 $10^{-13}$ J |
| $E_b/(\sigma^2_I+\sigma^2_G)$ | 17dB | 21dB | 12dB | 30dB |
| $\sigma^2_I/\sigma^2_G$ | 19 | 6 | 64 | 0.1 |
| FCC | | | | |
| $\sigma^2_I$ | 1.3 $10^{-10}$ J | 7.3 $10^{-12}$ J | 6.6 $10^{-10}$ J | 9.6 $10^{-14}$ J |
| $E_b/(\sigma^2_I+\sigma^2_G)$ | 16 dB | 29 dB | 9.3dB | 47dB |
| $\sigma^2_I/\sigma^2_G$ | 0.2 $10^6$ | 1.3 $10^4$ | 0.9 $10^6$ | 1.7 $10^2$ |

Table 3
**Summary of the measured impuilse noise data**

| Parameters | Average | Highest probability | Worst (10%) | Best( 10%) |
|---|---|---|---|---|
| Pulse duration | 36 μs | 6 μs | > 82 μs | < 2.1 μs |
| Interarrival time | 127 μs | 25 μs | > 14 μs | < 270 μs |
| A | 0.28 | 0.24 | 1 | 0.008 |

The average burst length is 14 symbols for CENELEC and 43 symbols for FCC.

III. IMPULSE NOISE CHANNEL MEASUREMENTS

Impulse energy, - time duration and - inter arrival time are used to estimate the influence of the impulse noise on OFDM and single carrier modulation schemes like S-FSK and BPSK. In Figure 1 and Tables 2-3 we give the results of our measurements for the CENELEC and FCC band, respectively.

From Table 2 and Table 3, we can observe, that impulse noise can have a severe impact on the performance of the OFDM modulation. We have to note that we did not include the attenuation of the signal (up to 100 dB per km). Table 2 indicates that reliable transmission is possible, but can become critical. Improvements can be obtained by using error correcting codes or a repeater, [9,10].

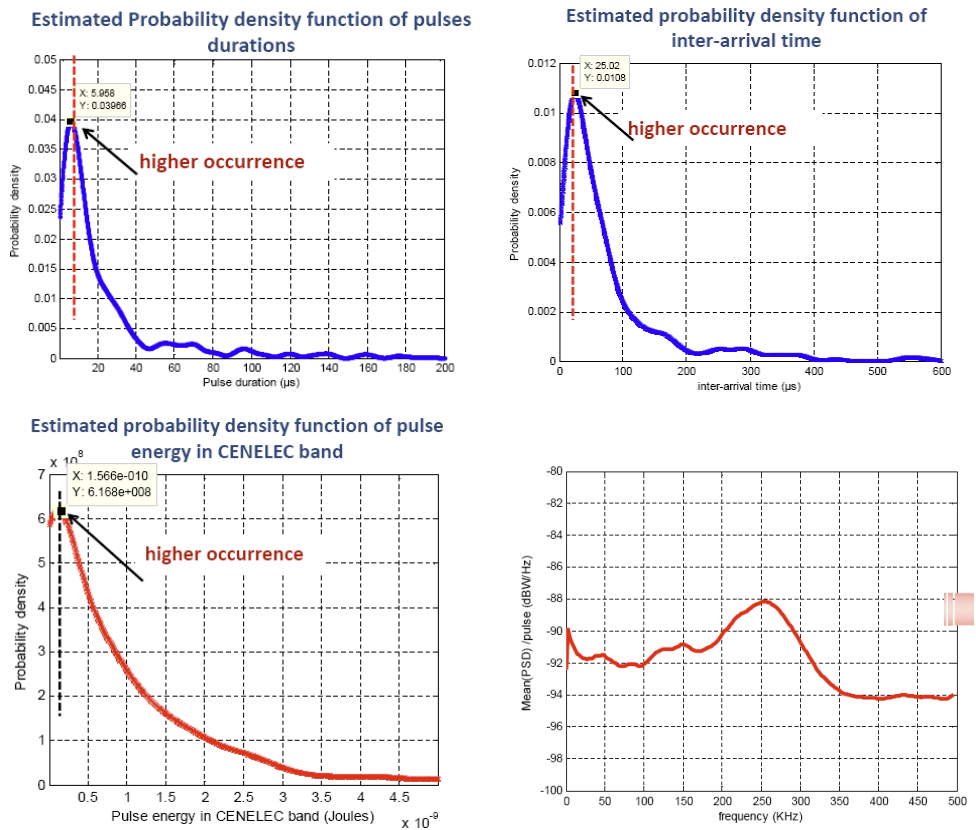

Fig.1. Properties of the impulse noise used in the calculations

IV. NOISE MITIGATION

Impulse noise mitigation in OFDM can be done in several ways. We summarize a few methods from literature:
1. *Nulling* [7]. At the receiver side a threshold is used to put the input signal to zero (null) when an impulse is detected. In this way the FFT output impulse noise variance is expected to be reduced from $\sigma^2_I$ to $A E_b$ when all impulses are detected. Additional thresholds can be used to improve the performance, [5]. Note that we consider the uncoded performance.
2. *Compressed sensing* [3,4]. The non-used carriers at the transmitter side can be used to estimate the noise at the receiver side.
3. *Iterative detection* [1,4,5]. An iterative receiver detects the impulse noise and uses this knowledge in an iterative way to estimate the transmitted OFDM frame.

Most of the methods assume full knowledge of the received signal level and noise parameters. In practice, this complicates the receiver and it might be difficult to realize the theoretical expectations.

## V. PERIODIC IMPULSE NOISE

Periodic impulse noise in the time domain gives rise to periodical impulse noise in the frequency domain. In Figure 3 we give the frequency domain output of the FFT for a particular periodic impulse noise input. Since the period of the FFT and the period of the impulse noise do not match, we have *leaking* contributions to neighboring positions. One way to reduce this effect is to randomize the transmitted samples in position and phase [6,11], see Figure 2. The resulting received samples can then be handled with methods as given in section IV. Of course, this results in more complexity at transmitter - and receiver side. If no randomization is used, then serious disturbances can occur in the frequency domain. Groups of sub-carriers might be involved and thus, convolutional codes must be followed by interleavers (more complexity). As an alternative, short block codes or codes correcting symbol errors can be used [6, 12, 13].

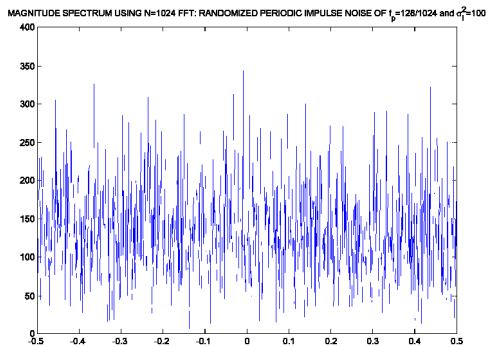 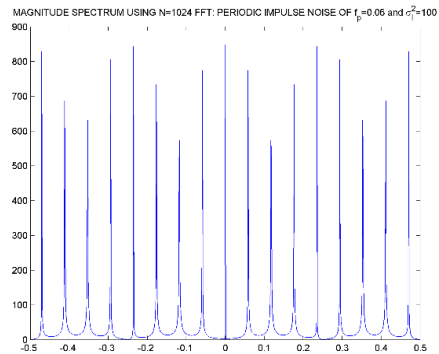

Fig. 2. Effect of randomization     Fig. 3. Frequency domain output for time periodic input

## VI. CONCLUSION

Impulse noise can have serious impact on the performance of OFDM based transmission in CENELEC and FCC band. The used approach can be extended to in-home broadband PLC communications. The difference in Signal to Noise Ratio is important for the determination of the maximum distance between transmitter and receiver. The application of error correcting codes is necessary to achieve acceptable performance.

## ACKNOWLEDGMENT

The authors would like to thank Miss Sabine Jankowski from the University of Duisburg-Essen, Institute of Digital Signal Processing, for her assistance in the management of this research.

## REFERENCES

[1] J. Häring and A.J. Han Vinck, *"Iterative decoding of codes over complex numbers for impulsive noise channels,"* IEEE Transactions on Information Theory, Jan. 2003, pp. 1251-1260
[2] M. Zimmermann, K. Dostert., *"Analysis and modeling of impulsive noise in broad-band powerline communications,"* IEEE Tr. on *EMC,* vol.44, no.1, Feb 2002, pp. 249-258
[3] L. Lampe, "*Bursty Impulse Noise Detection by Compressed Sensing*," ISPLC 2011, Udine, Italy, April 2011.
[4] A. Mengi, A. J. H. Vinck, *"Successive impulsive noise suppression in OFDM,"* ISPLC 2010, Rio de Janeiro, March 2010, pp. 33-37.
[5] V.Papilaya and A.J. Han Vinck, *"Improving Performance of the MH-Iterative IN Mitigation Scheme in PLC Systems,"* IEEE Transactions on Power Delivery, vol.30, no.1, 2015, pp.138-143,
[6] T. Shongwe, A.J. Han Vinck, H.C. Ferreira, *"The effects of periodic impulsive noise on OFDM,"* ISPLC 2015, March 2015, pp.189-194
[7] S. V. Zhidov, *"Impulse noise suppression in OFDM based communication systems,"* IEEE Trans.Cons. Elect. vol. 49, 2003, pp. 944-948
[8] J.A. Cortes, L. Dıez, F.J. Canete, and J.J. Sanchez-Martınez, *"Analysis of the Indoor Broadband Power-Line Noise Scenario,"* IEEE Transactions on EMC, Dec. 2010, pp. 849-858
[9] A.J. Han Vinck, *"Coding Concepts and Reed-Solomon Codes,"* ISBN 978-3-9813030-6-3, Inst. For Experimental Mathematics, Essen, Germany
[10] L. Lampe and A.J. Han Vinck, *"On Cooperative Coding for Narrowband PLC Networks,* " AEÜ, (AEUE ) (2011), doi:10.1016/j.aeue.2011
[11] J. Lin and B. Evans, *"Non-parametric mitigation of periodic impulsive noise in narrowband powerline communications,"* DOI: 10.1109 /GLOCOM.2013.6831528
[12] Shongwe, T.; Han Vinck, A.J., *"Broadband and Narrow-band Noise Modelling in Powerline Communications,"* Wiley Encyclopedia of Electrical and Electronics Engineering, to be published.
[13] T. Shongwe, A.J. Han Vinck, H.C Ferreira, *"The Effects of Periodic Impulsive Noise on OFDM,"* IEEE International Symposium on Powerline Communications and its Applications, March, 2015, Austin, USA (ISPC 2015), pp. 29-31